\documentclass[12pt]{article}
 
\def\cdate{{September 10, 2008}} 

\usepackage{
txfonts,
bm
} 
\usepackage[american]{babel} 

\def\II{{\mathbb I}}

\def\tr{\mathrm{ tr\,}} 
\def\Tr{\mathrm{ Tr\,}}

\def\be{\begin{equation}} 
\def\ee{\end{equation}} 
\def\bea{\begin{eqnarray}} 
\def\eea{\end{eqnarray}} 
\def\bed{\begin{definition}{\ }} 
\def\eed{\end{definition}} 
\def\bd{\begin{description}} 
\def\ed{\end{description}} 
\def\bc{\begin{center}} 
\def\ec{\end{center}}

\newtheorem{definition}{Definition} 
 
\def\sideremark#1{\ifvmode\leavevmode\fi\vadjust{\vbox to0pt{\vss 
\hbox to 0pt{\hskip\hsize\hskip1em 
\vbox{\hsize2cm\tiny\raggedright\pretolerance10000 
\noindent #1\hfill}\hss}\vbox to8pt{\vfil}\vss}}} 
 
\makeatletter 
\def\timenow{ 
\@tempcnta=\time \divide\@tempcnta by 60 
\number\@tempcnta:\multiply \@tempcnta by 60 \@tempcntb=\time 
\advance\@tempcntb by -\@tempcnta \ifnum\@tempcntb <10 
0\number\@tempcntb\else\number\@tempcntb\fi} 
\newcounter{outputpage} 
\renewcommand{\@oddhead} 
{\stepcounter{outputpage}\hfill\hfill\theoutputpage} 
\renewcommand{\@evenhead} 
{\stepcounter{outputpage}\hfill\hfill\theoutputpage} 
\renewcommand{\@oddfoot} 
{\vbox{ 
\vspace{3pt} \hfil {\scriptsize\textit{ 
\hfill\hfill
}} 
\hfil }} 
\renewcommand{\@evenfoot} 
{\vbox{ 
\vspace{3pt} \hfil {\scriptsize\textit{ 
\hfill\hfill
}} 
\hfil }} 
\makeatother 

\begin{document} 
 
\begin{titlepage} 
\thispagestyle{empty} \null 
 
\hspace*{50truemm}{\hrulefill}\par\vskip-4truemm\par 
\hspace*{50truemm}{\hrulefill}\par\vskip5mm\par 
\hspace*{50truemm}{{\large\sc New Mexico Tech {\rm 
(\cdate)}}}\vskip4mm\par 
\hspace*{50truemm}{\hrulefill}\par\vskip-4truemm\par 
\hspace*{50truemm}{\hrulefill} 
\par 
\bigskip 
\bigskip 
\par 
\par 
\vspace{1cm} 
\centerline{\huge\bf Kinematics in Matrix Gravity} 
\bigskip 
\bigskip 
\centerline{\Large\bf Ivan G. Avramidi and Guglielmo Fucci} 
\bigskip 
\centerline{\it New Mexico Institute of Mining and Technology} 
\centerline{\it Socorro, NM 87801, USA} 
\centerline{\it E-mail: 
iavramid@nmt.edu, gfucci@nmt.edu} 
\bigskip 
\medskip 
\vfill 
 
{\narrower 
\par 
We develop the kinematics in Matrix Gravity, which is a modified theory 
of gravity obtained by a non-commutative deformation of General 
Relativity. In this model the usual interpretation of gravity as
Riemannian  geometry is replaced by a new kind of geometry, which is
equivalent  to a collection of Finsler geometries with several Finsler
metrics depending  both on the position and on the velocity. As a result
the Riemannian geodesic  flow is replaced by a collection of Finsler
flows. This naturally leads to a model in which  a particle is described
by several mass parameters. 
If these mass parameters are different then the
equivalence principle is violated. In the non-relativistic limit this
also leads to corrections to the Newton's  gravitational potential. We
find the first and second order corrections to the usual Riemannian
geodesic flow and evaluate the anomalous nongeodesic  acceleration in a
particular case of static spherically symmetric background.

\par} 

\vfill 

 
\end{titlepage} 
 
\section{Introduction} 
\setcounter{equation}{0} 

Gravity is one of the most universal physical phenomenon. It is this universality
that leads to a successful geometric interpretation of gravity in terms of Riemannian
geometry in General Relativity. General Relativity is widely accepted as a pretty good
approximation to the physical reality at large range of scales. 

We would like to make two points here.
First of all, the experimental evidence points to the fact that all matter exhibits quantum behavior at microscopic scales. Thus, it is 
generally believed
that the classical
general relativistic description of gravity is inadequate at short distances
due to quantum fluctuations.
However, despite the enormous efforts to unify
gravity and quantum mechanics during the last several
decades we still do not have a consistent theory of quantum gravity. 
There are, of course, some promising approaches, like the string theory, 
loop gravity and non-commutative geometry. But, at the time, none of them
provides a complete consistent theory that can be verified by existing or 
realistic future experiments.

Secondly, in the last decade or so it became more and more evident that there 
might be a few problems  in 
the {\it classical
domain} as well. In addition to the old problem of gravitational singularities
in General Relativity these {\it gravitational anomalies} include such effects
as dark matter, dark energy, Pioneer anomaly, flyby anomaly, and others \cite{laemmerzahl06}. They 
might signal to {\it new physics} not only at the Planckian scales but 
at very large (galactic) scales as well. 

This suggests that General Relativity, that works perfectly well at macroscopic
scales, should be {\it modified (or deformed) 
both at microscopic and at galactic (or cosmological)
scales} (or, in the language of high energy physics, both in the ultraviolet 
and the infrared). It is very intriguing to imagine that these effects 
(that is, the quantum origin of gravity and gravitational anomalies at large 
scales) could be somehow related. 
Of course, this modification should be done in such a way that at the usual
distances the usual General Relativity is recovered. This condition puts some
constraints (experimental bounds) on the deformation parameters; in the case of non-commutative field theory such bounds on the non-commutativity parameter were
obtained in \cite{carroll01}.

In this paper we investigate the motion of test particles in an extended 
theory of gravity, called Matrix Gravity, proposed in a series of 
recent papers 
\cite{avramidi03,avramidi04a,avramidi04b}. 
The motivation for such a deformation of General Relativity is explained 
in detail in \cite{avramidi04a}. 
The very basic physical concepts are the notions of event and the spacetime.
An event is a collection of variables that specifies the location of a point
in space at a certain time. To assign a time to each point in space one needs
to place clocks at every point (say on a lattice in space) and to
synchronize these clocks. Once the position of the clocks is fixed the only way
to synchronize the clocks is by transmitting the information from a fixed point
(say, the origin of the coordinate system in space) to all other points. This can be
done by sending a signal through space from one point to another. Therefore,
the synchronization procedure depends on the propagation of the signal through
space, and, as a result, on the properties of the space it propagates through, 
in particular, on the presence of any physical background fields in space.
The propagation of signals is described by a wave equation (a hyperbolic 
partial differential equation of second order). Therefore, the propagation
of a signal depends on the matrix of the coefficients (a symmetric 2-tensor) $g^{\mu\nu}(x)$ of the second 
derivatives in the wave equation which must be non-degenerate and
have the signature $(-+\dots+)$. This matrix  
can be interpreted as a pseudo-Riemannian metric, which defines the geodesic flow,
the curvature and, finally, the Einstein equations of General Relativity
(for more details, see \cite{avramidi04a}).

The picture described above applies to the propagation of light, which is
described by a single wave equation. However, now we know that at microscopic scales
there are other fields that could be used to transmit a signal.
In particular, the propagation of a multiplet of $N$ gauge fields is described
not by a single wave equation but by a {\it hyperbolic system} 
of second order partial differential equations. The coefficients at the second derivatives of such a system are not given by just a 2-tensor like $g^{\mu\nu}(x)$
but by a $N\times N$ {\it matrix-valued} symmetric $2$-tensor $a^{\mu\nu}(x)$.
If $a^{\mu\nu}$ does not factorize as $a^{\mu\nu}\ne E g^{\mu\nu}$, where $E$
is some non-degenerate matrix, then there is no geometric
interpretation of this hyperbolic system in terms of a single Riemannian metric.
Instead, we obtain a new kind of geometry that we call Matrix Geometry, which is 
equivalent to a collection of Finsler geometries. In this theory, 
instead of a single Riemannian geodesic flow, there is a system of $N$ Finsler
geodesic flows. Moreover, a gravitating particle is described not by one mass
parameter but by $N$ mass parameters (which could be different). 
Note that because the tensor $a^{\mu\nu}$ is matrix-valued, various components of this
tensor do not commute, that is, $[a^{\mu\nu}, a^{\alpha\beta}]\ne 0$. In this sense
such geometry may be also called {\it non-commutative Riemannian geometry}.
In the commutative limit, $a^{\mu\nu}\to g^{\mu\nu}$ and we recover the
standard Riemannian geometry with all its ingredients. Only the 
total mass of a gravitating particle is observed. 
For more details and discussions see 
\cite{avramidi04a,avramidi04b}. 
 
As we outlined above, Matrix Gravity is a non-commutative
modification of the standard 
General Relativity in which the metric tensor $g^{\mu\nu}$ 
is replaced by a Hermitian $N\times N$ 
matrix-valued symmetric two-tensor
\be 
a^{\mu\nu}=g^{\mu\nu}\II+\varkappa h^{\mu\nu}\,, 
\label{1x} 
\ee 
where $\II$ is the identity matrix, 
$h^{\mu\nu}$ is a Hermitian 
matrix-valued traceless symmetric tensor, i.e. 
\be 
g^{\mu\nu}=\frac{1}{N}\tr a^{\mu\nu}\,,\qquad
\tr h^{\mu\nu}=0\,, 
\ee 
and $\varkappa$ is a  deformation parameter.

The dynamics of the tensor field $a^{\mu\nu}$ is described by a 
diffeomorphism invariant action, 
\be
S(a)=\int dx\; {\cal L}(a,\partial a)\,,
\ee
where $dx$ is the standard Lebesgue measure on the spacetime manifold and
${\cal L}(a,\partial a)$ is the Lagragian density.
Of course, as $\varkappa\to 0$ this action should reproduce the
usual Einstein-Hilbert action functional
\be
S(g)=\frac{1}{16 \pi G}\int dx\; g^{1/2}(R-2\Lambda)\,,
\ee
where $g_{\mu\nu}$ is a pseudo-Riemannian metric, $g=|\det g_{\mu\nu}|$,
$R$ is the scalar curvature of the metric $g$, 
$G$ is the Newton constant and $\Lambda$ is the cosmological constant.

The action of matrix gravity 
can be constructed in two different ways. One 
approach, developed in \cite{avramidi03,avramidi04a}, is to try to 
extend all standard concepts of differential geometry to the 
non-commutative setting and to construct a matrix-valued 
connection and a matrix-valued curvature.  

The second approach, 
developed in \cite{avramidi04b}, is based on the spectral asymptotics
of a self-adjoint elliptic partial differential operator $L$ of second 
order with a positive definite leading 
symbol $\sigma_L(x,\xi)=a^{\mu\nu}(x)\xi_\mu\xi_\nu$.
It is well known that there is an asymptotic expansion as $t\to 0$
of the $L^2$-trace of the heat semigroup of the operator $L$
\be
\Tr_{L^2}\exp(-tL)\sim (4\pi t)^{-n/2}\sum_{k=0}^\infty t^k A_k\,,
\ee
where $A_k$ are spectral invariants of the operator $L$.
For the Laplace-Beltrami operator
$L=-g^{\mu\nu}\nabla_\mu\nabla_\nu$ these coefficients are well known,
and, it turns out that the Einstein-Hilbert action is
nothing but a linear combination of the first two
coefficients, that is,
\be
S(g)=\frac{1}{16 \pi G}\Bigl(6A_1-2\Lambda A_0\Bigr)\,.
\ee
Therefore, a similar functional, which is automatically 
diffeomorphism-invariant, can be constructed by computing the same 
heat kernel coefficients for a more general partial differential 
operator $L$ of non-Laplace type (for more details, see 
\cite{avramidi04b}). 

The field equations for the tensor 
$a^{\mu\nu}$, that we call non-commutative Einstein 
equations are obtained by varying the action with respect to 
$a^{\mu\nu}$; in the vacuum we have,
\be
\frac{\partial {\cal L}}{\partial a^{\alpha\beta}}
-\partial_\mu
\frac{\partial {\cal L}}{\partial a^{\alpha\beta}{}_{,\mu}}=0\,,
\ee
where $a^{\alpha\beta}{}_{,\mu}=\partial_\mu a^{\alpha\beta}$.

The action  has an additional new {\it global} gauge symmetry 
\be 
a^{\mu\nu}(x) \mapsto U a^{\mu\nu}(x) U^{-1}\,, 
\ee 
where $U$ is a 
constant unitary 
matrix (for more details, see the papers cited above). 
By the Noether theorem
this symmetry leads to the conserved 
currents (vector densities)
$$
{\cal J}^\mu=\left[a^{\alpha\beta},\;{\partial {\cal L}\over\partial 
(a^{\alpha\beta}{}_{,\mu})}\right]\,,
\qquad
\partial_\mu {\cal J}^\mu=0\,.
$$
In other words, this suggests the existence of
{\it new physical charges} 
$$
Q=\int\limits d\hat x\, {\cal J}^0\,,
$$
where $d\hat x$ denotes the integration over the space coordinates only.
These charges have purely noncommutative origin and vanish in the 
commutative limit.

One can easily localize this global symmetry by introducing the 
{\it local} gauge transformations 
\be 
a^{\mu\nu}(x)\mapsto U(x)a^{\mu\nu}(x)U^{-1}(x)\,, 
\ee 
where $U(x)=\exp \omega(x)$, and $\omega$ is an anti-Hermitian
matrix-valued function of compact 
support, i.e. vanishing at infinity, and a new Yang-Mills field ${\cal 
B}_\mu$ that transforms as one-form under diffeomorphisms and as a 
connection under the gauge transformation. 
All the geometric structures, including the connection coefficients, the 
curvature etc, become covariant under the local gauge transformations if 
one simply replaces the partial derivatives  $\partial_\mu$ by 
$\partial_\mu+{\cal B}_\mu$. This leads to a gauged version of the above 
functionals (for more details see 
\cite{avramidi03,avramidi04a,avramidi04b}). 
 
This model may be viewed as a ``noncommutative deformation'' of Einstein 
gravity (coupled to a Yang-Mills model  in the gauged version), which 
describes, in the weak deformation limit, as $\varkappa\to 0$, General 
Relativity, Yang-Mills fields (in the gauged version), and a multiplet 
of self-interacting massive two-tensor fields of spin $2$ that interact 
also with gravity and the Yang-Mills fields. 
One should stress here that $\varkappa$ is a {\it formal} parameter
that does not have a particular physical value; it is just a tool
to develop the perturbation theory in $h^{\mu\nu}$.
At the end of the derivation we can just set $\varkappa=1$. 
What is measured and describes the extent of the non-commutative deformation
is the tensor $h^{\mu\nu}(x)$ and its derivatives, which can be parametrized
by he invariants of this tensor like
$g_{\mu\alpha}g_{\nu\beta} \frac{1}{N}\tr h^{\mu\nu}h^{\alpha\beta}$.

Our approach should be contrasted with the non-commutative extensions of 
gravity on non-commutative spaces with non-commutative coordinates
\be
[x^\mu,x^\nu]=\theta^{\mu\nu}\,,
\label{110iga}
\ee
where $\theta^{\mu\nu}$ is a constant anti-symmetric matrix, and 
the Moyal product
\be
f(x)\star g(x)=\exp\left(\frac{i}{2}\theta^{\mu\nu}
\frac{\partial}{\partial y^\mu}\frac{\partial}{\partial z^\nu}\right)
f(x+y)g(x+z)\Big|_{y=z=0}\,.
\ee
This approach immediately rasies the question on the nature of the coordinates
$x^\mu$. The non-commutativity condition can only be true for some priviledged
coordinates, like Cartesian (or inertial) coordinates.
The condition that the matrix $\theta^{\mu\nu}$ is constant, that is,
\be
\partial_\alpha\theta^{\mu\nu}=0\,,
\ee
is not covariant. It breaks the diffeomorphism invariance of the theory,
and, as the result, Lorentz invariance. Thus any such theory cannot be diffeomorphism
invariant. One could try to replace it by the 
covariant condition
\be
\nabla_\alpha\theta^{\mu\nu}=0\,,
\ee
where $\nabla_\alpha$ are covariant derivatives with respect to some 
background metric. Then the matrix $\theta^{\mu\nu}$ could be viewed  as
a covariantly constant antisymmetric $2$-tensor. However, the integrability 
conditions for this equation lead to very strong algebraic constraints on the
curvature of the metric. Thus,
the breaking of the diffeomorphism invariance (and as a result of Lorentz
invariance) is an unavoidable feature of this approach to non-commutative gravity.
Therefore, such theories can be ruled out by very restrictive experimental bounds
on the non-commutativity parameter \cite{carroll01}.

By contrast, the status of diffeomorphism invariance (and the Lorentz invariance)
in Matrix Gravity is exactly the same as in General Relativity, namely, both
theories are diffeomorphism invariant, so, there are no preferred coordinates, and
a condition like (\ref{110iga}) is impossible. In our model {\it it is not the coordinates that do not commute, but the metric!} 
Therefore, the recent strong experimental constraints on the violation of
Lorentz invariance do not apply to Matrix Gravity.
It is rather the {\it violation of the Equivalence Principle} that is critical for
Matrix Gravity. This feature could be used for an experimental test of the
theory in the future.

One should also mention the relation of our approach to so called 
``analog models of gravity''. In particular, the analysis in \cite{barcelo02}
is surprisingly similar to the analysis of our papers \cite{avramidi03,avramidi04a}.
The authors of \cite{barcelo02} 
consider a hyperbolic system of second order partial differential equations,
the corresponding Hamilton-Jacobi equations and the Hamiltonian system
as we did in \cite{avramidi03,avramidi04a}. 
In fact, their ${\bf f}^{\mu\nu}$ is equivalent to our matrix-valued
tensor $a^{\mu\nu}$, 
However, their goal was very different---they impose the commutativity conditions
on ${\bf f}^{\mu\nu}$ (eq. (44)) to enforce a unique effective metric for the
compatibility with the Equivalence Principle.
They barely mention the general geometric interpretation in terms
of Finsler geometries as it ``does not seem to be immediately relevant for either particle physics or gravitation'' 
The motivation of the authors of
\cite{barcelo02} is also very different from our approach. Their idea is that gravity
is not fundamental so that the effective metric simply 
reflects the properties of an underlying physics (such as fluid meachnics and 
condensed matter theory). They just need to have enough fields to be able to parametrize
an arbitrary effective metric. In our approach, the matrix-valued field 
$a^{\mu\nu}$ is fundamental; it is: i) non-commutative and ii) dynamical.

The main goal of the present paper is to investigate the motion of
test particles in a simple model of matrix gravity and study the 
non-geodesic corrections to general relativity.
 
The outline of this work is as follows. In Sect. 2.  we develop the
kinematics in Matrix Gravity. In Sect. 3. we compute the first and
second order  non-commutative corrections to the usual Riemannian
geodesic flow. In Sect. 4 we find a static spherically  symmetric
solution of the dynamical equations of Matrix Gravity in a  particular
case of commutative $2\times 2$ matrices. In Sect. 5 we  evaluate the
anomalous acceleration of test particles in this background. In Sect. 6
we discuss our results.
 
\section{Kinematics in Matrix Gravity} 
\setcounter{equation}{0} 
 
\subsection{Riemannian Geometry} 
 
Let us recall how the geodesic motion appears in General 
Relativity, that is, in Riemannian  geometry (for more details, 
see \cite{avramidi04a}).  First of all, let 
\be 
F(x,\xi)=\sqrt{-|\xi|^2}\,, 
\ee 
where 
$\xi_\mu$ is a non-vanishing cotangent vector at the point $x$,
and $|\xi|^2=g^{\mu\nu}(x)\xi_\mu\xi_\nu$
(recall that the signature of our metric is $(-+\dots+)$). 
Obviously, this is a homogeneous function of $\xi$ of degree $1$, 
that is, 
\be 
F(x,\lambda\xi)=\lambda F(x,\xi)\,. 
\ee 
Let 
\be 
H(x,\xi)=-\frac{1}{2}F^2(x,\xi) 
=\frac{1}{2}|\xi|^2\,. 
\ee 
This is, of course, 
a homogeneous polynomial of $\xi_\mu$ of order $2$, and, 
therefore,  the Riemannian metric can be recovered by 
\be 
g^{\mu\nu}(x)= 
\frac{\partial^2}{\partial\xi_\mu\partial\xi_\nu}H(x,\xi)\,. 
\ee 
Now, let us consider a Hamiltonian system with the Hamiltonian 
$H(x,\xi)$ 
\bea 
\frac{dx^\mu}{dt}&=& \frac{\partial 
H(x,\xi)}{\partial \xi_\mu} 
=g^{\mu\nu}(x)\xi_\nu\,,\\[12pt] 
\frac{d\xi_\mu}{dt}&=& \frac{\partial H(x,\xi)}{\partial x^\mu} = 
-\frac{1}{2}\partial_\mu g^{\alpha\beta}(x) \xi_\alpha\xi_\beta\,. 
\eea 
The trajectories of this Hamiltonian system are, then, 
nothing but the geodesics of the metric $g_{\mu\nu}$. Of course, 
the Hamiltonian is conserved, that is, 
\be 
g^{\mu\nu}(x(t))\xi_\mu(t)\xi_\nu(t) =-E\,, 
\ee 
where $E$ is a constant 
parameter.

\subsection{Finsler Geometry} 
 
As it is explained in \cite{avramidi04a,avramidi04b} Matrix 
Gravity is closely related to {\it Finsler geometry} \cite{rund59} 
rather than Riemannian geometry. In this section we 
follow the description of Finsler geometry outlined in 
\cite{rund59}. 
To avoid confusion we should note that we present
it in a slightly modified equivalent form, namely, we start with the
Finsler function in the cotangent bundle rather than in the 
tangent bundle.

Finsler geometry is defined by a 
Finsler function $F(x,\xi)$ which is a homogeneous function of 
$\xi_\mu$ of degree $1$ and the Hamiltonian 
\be\label{3} 
H(x,\xi)=-\frac{1}{2}F^2(x,\xi)\,. 
\ee 
Such Hamiltonian is still 
a homogeneous function of $\xi_\mu$ of degree $2$, that is, 
\be 
\xi_\mu\frac{\partial}{\partial \xi_\mu}H(x,\xi)=2H(x,\xi)\,, 
\ee 
but {\it not necessarily a polynomial} in $\xi_\mu$! 
 
Now, we define a tangent vector $u$ by 
\be 
u^\mu=\frac{\partial}{\partial\xi_\mu}H(x,\xi)\,, 
\ee 
and the \emph{Finsler metric} 
\be 
\label{6} 
G^{\mu\nu}(x,\xi)=\frac{\partial^2} 
{\partial\xi_\mu\partial\xi_\nu }H(x,\xi)\,. 
\ee 
 
The difference with the Riemannian metric is, obviously, that the 
Finsler metric does depend on $\xi_\mu$, more precisely, it is a 
homogeneous function of $\xi_\mu$ of degree $0$, i.e. 
\be 
G^{\mu\nu}(x,\lambda\xi)=G^{\mu\nu}(x,\xi)\,, 
\ee 
so that it 
depends only on the direction of the covector $\xi$ but not on its 
magnitude. This leads to a number of useful identities, in 
particular, 
\be 
H(x,\xi)=\frac{1}{2} 
G^{\mu\nu}(x,\xi)\xi_\mu\xi_\nu\,, 
\ee 
and 
\be 
u^\mu=G^{\mu\nu}(x,\xi)\xi_\nu\,. 
\ee 
 
Now, we can solve this equation for $\xi_\mu$ treating $u^\nu$ as 
independent variables to get 
\be 
\xi_\mu=G_{\mu\nu}(x,u)u^\nu\,, 
\ee 
where $G_{\mu\nu}$ is the inverse Finsler metric defined by 
\be 
G_{\mu\nu}(x,u) G^{\nu\alpha}(x,\xi)=\delta^\alpha_\mu\,. 
\ee 
By using the results obtained above we can express the Hamiltonian 
$H$ in terms of the vector $u^{\mu}$, more precisely we have 
\be 
H(x,\xi(x,u))=\frac{1}{2} G_{\mu\nu}(x,u)u^\mu u^\nu\,. 
\ee 
 
The derivatives of the Finsler metric obviously satisfy the 
identities 
\be 
\frac{\partial}{\partial 
\xi_\alpha}G^{\beta\gamma}(x,\xi) =\frac{\partial}{\partial 
\xi_\beta}G^{\gamma\alpha}(x,\xi) =\frac{\partial}{\partial 
\xi_\gamma}G^{\alpha\beta}(x,\xi)\,, 
\ee 
\be 
\xi_\mu\frac{\partial}{\partial \xi_\mu}G^{\nu\alpha}(x,\xi) = 
\xi_\mu\frac{\partial}{\partial \xi_\nu}G^{\mu\alpha}(x,\xi)= 0\,, 
\ee 
and, more generally, 
\be 
\xi_\mu \frac{\partial^k}{\partial 
\xi_{\nu_1} \dots\partial \xi_{\nu_k}}G^{\mu\alpha}(x,\xi) =0\,. 
\ee 
This means, in particular, that the following relations hold 
\be 
\frac{\partial u^\mu}{\partial 
\xi_\alpha}=G^{\mu\alpha}(x,\xi)\,, \qquad \frac{\partial 
\xi_\alpha}{\partial u^\mu}=G_{\mu\alpha}(x,u)\,. 
\ee 
 
It is easy to see that the metric $G_{\mu\nu}(x,u)$ is a 
homogeneous function of $u$ of degree $0$, that is, 
\be 
u^\mu\frac{\partial }{\partial u^\mu}G_{\nu\alpha}(x,u)=0\,, 
\ee 
and, therefore, $H(x,\xi(x,u))$ is a homogeneous function of $u$ 
of degree $2$. This leads to the identities 
\be 
\xi_\mu=\frac{1}{2} 
{\partial \over\partial u^\mu}H(x,\xi(x,u))\,, 
\ee 
\be 
G_{\mu\nu}(x,u)={1\over 2} {\partial^2 \over\partial u^\mu\partial 
u^\nu }H(x,\xi(x,u))\,. 
\ee 
 
Finally, this enables one to define the Finsler interval 
\be 
ds^2=G_{\mu\nu}(x,\dot x)dx^\mu dx^\nu\,, 
\ee so that 
\be 
d\tau=\sqrt{-ds^2}= \sqrt{-G_{\mu\nu}(x,\dot x)\dot x^\mu\dot 
x^\nu}\; dt =F(x,\xi(x,\dot x))dt\,, \label{33f} 
\ee 
where 
\be 
\dot x^\mu=\frac{dx^\mu}{dt}\,,\qquad \xi_\mu=G_{\mu\nu}(x,\dot 
x)\dot x^\nu\,. 
\ee 
By treating $H(x,\xi)$ as a Hamiltonian we 
obtain a system of first order ordinary differential equations 
\bea 
\frac{dx^\mu}{dt}&=& \frac{\partial H(x,\xi)}{\partial 
\xi_\mu} 
\,,\\[12pt] 
\frac{d\xi_\mu}{dt}&=&- \frac{\partial H(x,\xi)}{\partial 
x^\mu}\,. 
\eea 
The trajectories of this Hamiltonian system 
naturally replace the geodesics in Riemannian geometry.  Again, as 
in the Riemannian case, the Hamiltonian is conserved along the 
integral trajectories 
\be 
H(x(t),\xi(t))=-E\,. 
\ee 
Of course, in 
the particular case, when the Hamiltonian is equal to 
$H(x,\xi)=\frac{1}{2}|\xi|^2$,  all the 
constructions derived above reduce to the standard structure of Riemannian 
geometry.

\subsection{Matrix Gravity} 
 
The kinematics in Matrix Gravity is defined as follows. In 
complete analogy with the above discussion we consider the matrix 
\be 
A(x,\xi)=a^{\mu\nu}(x)\xi_\mu\xi_\nu\,, 
\ee 
where $a^{\mu\nu}$ is the matrix-valued metric (\ref{1x}).
As we mentioned in the introdution this expression has been already
encountered in physics, in particular, in \cite{barcelo02} it is shown 
that it is the most general structure describing ``analog models'' for gravity.

This is a 
Hermitian matrix, so it has real eigenvalues $h_{i}(x,\xi)$, 
$i=1,2,\dots, N$. We consider a generic case when the eigenvalues 
are simple. We note that the eigenvalues $h_{i}(x,\xi)$ are 
homogeneous functions (but not polynomials!) of $\xi$ of degree 
$2$. Thus, each one of them, more precisely $\sqrt{-h_i(x,\xi)}$, 
can serve as a Finsler function. In other words, we obtain $N$ 
different Finsler functions, and, therefore, $N$ different Finsler 
metrics. Thus, quite naturally, instead of a single Riemannian 
metric and a unique Riemannian geodesic flow there appears $N$ 
Finsler metrics and $N$ corresponding flows. In some sense, the 
noncommutativity leads to a ``splitting'' of a single geodesic to 
a system of close trajectories. 
 
Now, to define a unique Finsler metric
we need to define a unique Hamiltonian, which is a homogeneous
function of the momenta of degree $2$. It is defined in terms of the 
Finsler function as in (\ref{3}) 
which is a homogeneous function of the momenta of degree
$1$.
To define a unique Finsler function we can proceed as follows. Let 
$\mu_i$, $i=1,\dots, N$, be some dimensionless real parameters 
such that 
\be\label{14} 
\sum_{i=1}^N \mu_i=1\,, 
\ee 
so that there 
are $(N-1)$ independent parameters. Then we can define the Finsler 
function by 
\be 
F(x,\xi)=\sum_{i=1}^N \mu_i \sqrt{-h_i(x,\xi)}\,. 
\label{233xx}
\ee 
Notice that,  in the commutative limit, as $\varkappa\to 0$ 
and $a^{\mu\nu}=g^{\mu\nu}\II$,  all eigenvalues of the matrix 
$A(x,\xi)$ degenerate to the same  value, 
$h_i(x,\xi)=|\xi|^2$, and, hence, the Finsler 
function becomes $F(x,\xi)=\sqrt{-|\xi|^2}$. 
In this case the Finsler flow degenerates to the usual Riemannian 
geodesic flow. 
 
Next, we define the Hamiltonian according to eq. (\ref{3})
\bea 
H(x,\xi)&=&
-\frac{1}{2} 
\left(\sum_{i=1}^N \mu_i \sqrt{-h_i(x,\xi)}\right)^2 
\nonumber\\ 
&=& 
\frac{1}{2}\sum_{i=1}^N \mu_i^2 h_i(x,\xi) 
-\sum_{1\le i<j\le N}\mu_i\mu_j \sqrt{h_i(x,\xi)h_j(x,\xi)}\,. 
\label{234xx}
\eea 
 
In a particular case, when all parameters $\mu_i$ are equal, i.e. 
$\mu_i=1/N$, the Finsler function reduces to 
\be 
F(x,\xi)=\frac{1}{N}\sum_{i=1}^N \sqrt{-h_i(x,\xi)} 
=\frac{1}{N}\;\tr \sqrt{-A(x,\xi)}\,. 
\label{235xx}
\ee 
By using the 
decomposition of the matrix-valued metric $a^{\mu\nu}$ as in 
(\ref{1x}) one can see that 
\be 
\frac{1}{N}\tr 
A(x,\xi)=|\xi|^2\,, 
\ee 
and, therefore, 
\be 
\frac{1}{N}\sum_{i=1}^N h_i(x,\xi)=|\xi|^2\,. 
\ee 
Thus, we conclude that in this particular case 
\be H(x,\xi) 
=\frac{1}{N}\left( \frac{1}{2} |\xi|^2 
-\frac{1}{N} \sum_{1\le i<j\le N} 
\sqrt{h_i(x,\xi)h_j(x,\xi)}\right)\,. 
\ee 
 
It is difficult to give a general physical picture of these models since
the Hamiltonian is non-polynomial in the momenta. Hamiltonian systems with
homogeneous Hamiltonians have not been studied as thoroughly as the
usual systems with quadratic Hamiltonians and a potential.

\subsection{Kinematics} 
 
The problem is, now, how to use these mathematical tools to 
describe the motion of physical massive test particles in Matrix 
Gravity. The motion of a massive
particle in the gravitational field is 
determined 
in General Relativity by the action which is proportional to the 
interval, that, is,
\be 
S_{\rm particle} =-\int\limits_{P_1}^{P_2} 
m\sqrt{-g_{\mu\nu}(x)dx^\mu dx^\nu} =-\int\limits_{t_1}^{t_2} m 
\sqrt{-|\dot x|^2}dt\,, 
\ee 
where 
$m$ is the mass of the particle, 
$P_1$ and $P_2$ are the initial and the final position of the 
particle in the spacetime,
$t$ is a parameter, $t_1$ and $t_2$ are the initial and the final 
values, $\dot x^\mu=\frac{dx^\mu}{dt}$ and 
$|\dot x|^2=g_{\mu\nu}(x)\dot 
x^\mu\dot x^\nu\,.$ 
This action is, of course, 
reparametrization-invariant. So, as always, there is a freedom of 
choosing the parameter $t$. We can always choose the parameter to be the 
{\it affine parameter}  such that $|\dot x|^2$ is constant, for example, 
if the parameter is the proper time $t=\tau$, then $|\dot x|^2=-1$. 
The Euler-Lagrange equations for this functional are, of course, 
\be 
\frac{D \dot x^\nu}{dt} 
=\frac{d^2 \dot x^\nu}{dt^2} 
+\Gamma^\nu{}_{\alpha\beta}(x)\dot x^\alpha\dot x^\beta 
=0\,, 
\ee 
where $\Gamma^\mu{}_{\alpha\beta}$ are the standard 
Christoffel symbols of the metric 
$g_{\mu\nu}$. 
Of course, the equivalence principle holds since these equations do not 
depend on the mass. 

In Matrix Gravity a particle is described
instead of one mass 
parameter $m$ by   $N$ different mass 
parameters 
\be 
m_i=m\mu_i\,, 
\ee 
where 
\be 
m=\sum_{i=1}^N m_i\,.
\ee 
The parameters
$m_i$  
describe the ``tendency'' for a particle to move along 
the trajectory determined by the corresponding Hamiltonian 
$h_i(x,\xi)$. 
 In the commutative limit we only observe the total mass $m$. 
 
We define the Finsler function $F(x,\xi)$ and the Hamiltonian 
$H(x,\xi)$ as in eqs. (\ref{233xx}) and (\ref{234xx}).
Then the action for a 
particle in the gravitational field has the form 
\be 
S_{\rm 
particle} =-\int\limits_{t_1}^{t_2} m F(x,\xi(x,\dot x))\;dt \,. 
\ee 
Thus, the Finsler function $F(x,\xi(x,\dot x))$ (with the 
covector $\xi_\mu$ expressed in terms of the tangent vector $\dot 
x^\mu$) plays the role of the Lagrangian. To study the role of 
non-commutative corrections, it is convenient to rewrite this 
action in the form that resembles the action in General 
Relativity. 
\be 
S_{\rm particle} 
=-\int\limits_{t_1}^{t_2} m_{\rm eff}(x,\dot x) \sqrt{-|\dot x|^2}dt\,, 
\ee 
with some 
{\it ``effective mass'' $m_{\rm eff}(x,\dot x)$ that depends 
on the location and on the velocity of the particle} 
\be 
m_{\rm eff}(x,\dot x) 
=\sum_{i=1}^N m_{i} 
\sqrt{\frac{h_i(x,\xi(\dot x))}{|\dot x|^2}}\,. 
\ee

This action is again reparametrization-invariant. Therefore, we 
can choose the natural arc-length parameter so that 
$F(x,\xi(x,\dot x))=1$. Then the equations of motion determined by 
the Euler-Lagrange equations have the same form 
\be 
\frac{d^2 
x^\mu}{dt^2}+\gamma^\mu{}_{\alpha\beta}(x,\dot x)\dot x^\alpha \dot 
x^\beta=0\,, 
\ee 
where $\gamma^\mu{}_{\alpha\beta}(x,\dot x)$ are 
the Finsler Christoffel coefficients defined by the equations that 
look identical to the usual equations but with the Finsler metric 
instead of the Riemannian metric, that is, 
\be\label{9} 
\gamma^\mu{}_{\alpha\beta}(x,\dot x) 
=\frac{1}{2}G^{\mu\nu}(x,\xi(x,\dot x)) 
\left(\frac{\partial}{\partial x^\alpha} G_{\nu\beta}(x,\dot x) 
+\frac{\partial}{\partial x^\beta} G_{\nu\alpha}(x,\dot x) 
-\frac{\partial}{\partial x^\nu} G_{\alpha\beta}(x,\dot x) 
\right)\,. 
\ee 
 
To study the role of non-commutative corrections it is convenient 
to rewrite these equations in a covariant form in the Riemannian 
language.  In 
the commutative limit, as $\varkappa\to 0$, we can expand all our 
constructions in power series in $\varkappa$ so that the 
non-perturbed quantities are the Riemannian ones. In particular, 
we have 
\be 
\gamma^\mu{}_{\alpha\beta}(x,\dot x)=\Gamma^\mu{}_{\alpha\beta}(x) 
+\theta^\mu{}_{\alpha\beta}(x,\dot x)\,, 
\ee 
where 
$\theta^\mu{}_{\alpha\beta}$ are some tensors of order $\varkappa$. 
Then the equations of motion can be written in the form 
\be 
\frac{D \dot x^\nu}{dt}=A^\nu_{\rm anom}(x,\dot x)\,, 
\ee 
where 
\be 
\frac{D \dot x^\nu}{dt} =\frac{d^2 
x^\mu}{dt^2}+\Gamma^\mu{}_{\alpha\beta}(x)\dot x^\alpha \dot x^\beta 
\ee 
and 
\be\label{12} 
A^\nu_{\rm anom}(x,\dot 
x)=-\theta^\nu{}_{\alpha\beta}(x,\dot x) \dot x^\alpha \dot 
x^\beta\,, 
\ee 
is the {\it anomalous nongeodesic acceleration}. 
 
\section{Perturbation Theory} 
 
We see that the motion of test particles in matrix Gravity is quite 
different from that of General Relativity. The most important difference 
is that particles exhibit a {\it non-geodesic motion}. In other words, 
there is no Riemannian metric such that particles move along the 
geodesics of that metric. It is this anomalous acceleration that we are 
going to study in this paper. 
 
In the commutative limit the action of a particle in Matrix Gravity 
reduces to the action of a particle in General  Relativity with the mass 
$m$ determined by the sum of all masses $m_i$.  In 
this paper we consider  two different cases. In the first case, that we 
call the {\it nonuniform model}, we assume that all mass parameters are 
different, and in the second case, that we call the {\it uniform model}, 
we discuss what happens if they are equal to each other. 
 
\subsection{Nonuniform Model: First Order in $\varkappa$} 
 
So, in this section we study the generic case when the parameters 
$\mu_i$ are different. 
As we already mentioned above, in this case the Finsler function 
$F(x,\xi)$ is given by (\ref{233xx}).
By using the decomposition (\ref{1x}) of the matrix-valued metric 
$a^{\mu\nu}$ we have 
\be 
A(x,\xi)=a^{\mu\nu}(x)\xi_\mu\xi_\nu 
=|\xi|^2\II+\varkappa h^{\mu\nu}(x)\xi_\mu\xi_\nu\,.
\ee 
Therefore, the 
eigenvalues of the matrix $A(x,\xi)$ are 
\be 
h_i(x,\xi)=|\xi|^2+\varkappa\lambda_i(x,\xi)\,, 
\ee 
where 
$\lambda_i(x,\xi)$ are the eigenvalues of the matrix 
$h^{\mu\nu}(x)\xi_\mu\xi_\nu$. In the first order in $\varkappa$ 
we get the Finsler function 
\be F(x,\xi)=\sqrt{-|\xi|^2}\left(1 
+\varkappa\frac{1}{2}\frac{P(x,\xi)}{|\xi|^2}\right) 
+O(\varkappa^2)\,, 
\ee 
and the Hamiltonian 
\bea 
H(x,\xi)&=&\frac{1}{2}|\xi|^2 +\varkappa\frac{1}{2}P(x,\xi) 
+O(\varkappa^2)\,, 
\nonumber\\ 
\eea 
where 
\be 
P(x,\xi)=\sum_{i=1}^N\mu_i\lambda_i(x,\xi)\,. 
\ee 
By using the fact
that $P(x,\xi)$ is a homogeneous function of $\xi$ of order 
$2$, we find the Finsler metric 
\be 
G^{\mu\nu}(x,\xi)=g^{\mu\nu}(x) 
+\varkappa q^{\mu\nu}(x,\xi) 
+O(\varkappa^2)\,, 
\ee 
and its inverse 
\be 
G_{\mu\nu}(x,u)=g_{\mu\nu}(x) 
-\varkappa q_{\mu\nu}(x,\xi(x,u)) 
+O(\varkappa^2)\,, 
\ee 
where 
\be 
q^{\mu\nu}(x,\xi) 
=\frac{1}{2}\frac{\partial^2}{\partial \xi_\mu\partial\xi_\nu} 
P(x,\xi)\,. 
\label{275xx} 
\ee 
Here the indices are raised and lowered with the Riemannian 
metric, and 
\be 
u^\mu(x,\xi)=G^{\mu\nu}(x,\xi)\xi_\nu\,,\qquad 
\xi_\mu(x,u)=G_{\mu\nu}(x,u)u^\nu\,. 
\ee 
Since $P(x,\xi)$ is a homogeneous function of $\xi$ of order $2$ 
we have 
\be 
P(x,\xi)=q^{\mu\nu}(x,\xi)\xi_\mu\xi_\nu\,. 
\ee 
Note that since $\tr h^{\mu\nu}=0$ the matrix $h^{\mu\nu}\xi_\mu\xi_\nu$ 
is traceless, which implies that the sum of its eigenvalues is equal to 
zero. Thus, in the uniform case, when all mass parameters $\mu_i$ are 
the same, the function $P(x,\xi)$ vanishes. In this case the effects of 
non-commutativity are of the second order in $\varkappa$; we study this 
case in the next section. 
 
We also note that 
\be 
|\xi|^2=|u|^2-2\varkappa P(x,\xi(x,u)) 
+O(\varkappa^2)\,. 
\ee 
Thus, our Lagrangian is 
\be 
F(x,\xi(x,\dot x))= 
\sqrt{-|\dot x|^2}\left(1 
-\varkappa \frac{1}{2} 
\frac{P(x,\xi(x,\dot x))}{|\dot x|^2} 
\right) 
+O(\varkappa^2)\,, 
\ee 
Finally, we compute the Christoffel symbols 
to obtain 
\be 
\theta^\mu{}_{\alpha\beta}(x,\dot x) 
=-\frac{1}{2}\varkappa g^{\mu\nu}\left( 
\nabla_\alpha q_{\beta\nu}(x,\dot x) 
+\nabla_\beta q_{\alpha\nu}(x,\dot x) 
-\nabla_\nu q_{\alpha\beta}(x,\dot x) 
\right) 
+O(\varkappa^2)\,, 
\ee 
and the covariant derivatives are defined with the Riemannian metric. 
 
Thus, the anomalous acceleration is 
\be 
A^\mu{}_{\rm anom}= 
\frac{\varkappa}{2} 
g^{\mu\nu}\left( 
2\nabla_\alpha q_{\beta\nu}(x,\dot x) 
-\nabla_\nu q_{\alpha\beta}(x,\dot x) 
\right)\dot x^\alpha \dot x^\beta 
+O(\varkappa^2)\,, 
\ee 
 
\subsection{Uniform Model: Second Order in $\varkappa$} 
 
So, in this section we will simply assume that all mass parameters 
are equal, that 
is, 
\be 
m_i=\frac{m}{N}\,. 
\ee 
In this case the Finsler function 
$F(x,\xi)$ is given by (\ref{235xx}).
By using the decomposition of the matrix-valued metric and 
the fact that $\tr h^{\mu\nu}=0$ we get 
the Finsler function 
 \be 
F(x,\xi)=\sqrt{-|\xi|^2}\left(1 
-\varkappa^2\frac{1}{8}S^{\mu\nu\alpha\beta}(x) 
\frac{\xi_\mu\xi_\nu\xi_\alpha\xi_\beta}{|\xi|^4}\right) 
+O(\varkappa^3)\,, 
\ee 
and the Hamiltonian 
\bea 
H(x,\xi)&=&\frac{1}{2}|\xi|^2\left(1 
-\varkappa^2\frac{1}{4}S^{\mu\nu\alpha\beta}(x) 
\frac{\xi_\mu\xi_\nu\xi_\alpha\xi_\beta}{|\xi|^4}\right) 
+O(\varkappa^3)\,, 
\eea 
where 
\be 
S^{\mu\nu\alpha\beta}=\frac{1}{N}\tr 
(h^{\mu\nu}h^{\alpha\beta})\,. 
\ee 
By using the above, we compute the 
Finsler metric 
\be 
G^{\mu\nu}(x,\xi)=g^{\mu\nu}(x) 
-\varkappa^2\frac{1}{4}S^{\mu\nu\alpha\beta}(x) 
\frac{\xi_\alpha\xi_\beta}{|\xi|^2} 
+O(\varkappa^3)\,, 
\ee 
and its inverse 
\be 
G_{\mu\nu}(x,u)=g_{\mu\nu}(x) 
+\varkappa^2\frac{1}{4}S_{\mu\nu\alpha\beta}(x) 
\frac{u^\alpha u^\beta}{|u|^2} 
+O(\varkappa^3)\,. 
\ee 
We also note that 
\be 
|\xi|^2=|u|^2+\varkappa^2\frac{1}{2}S_{\mu\nu\alpha\beta}(x) 
\frac{u^\mu u^\nu u^\alpha u^\beta}{|u|^2} 
+O(\varkappa^3)\,. 
\ee 
Thus, our Lagrangian is 
\be 
F(x,\xi(x,\dot x))= 
\sqrt{-|\dot x|^2}\left(1 
+\varkappa^2\frac{1}{8}S_{\mu\nu\alpha\beta}(x) 
\frac{\dot x^\mu \dot x^\nu \dot x^\alpha \dot x^\beta}{|\dot x|^4}\right) 
+O(\varkappa^3)\,, 
\ee 
Finally, we 
compute the Christoffel symbols to obtain 
\be 
\theta^\mu{}_{\alpha\beta}(x,\dot x) 
=\varkappa^2\frac{1}{8}g^{\mu\nu}\left( 
\nabla_\alpha S_{\beta\nu\rho\sigma} 
+\nabla_\beta S_{\alpha\nu\rho\sigma} 
-\nabla_\nu S_{\alpha\beta\rho\sigma} 
\right)\frac{\dot x^\rho\dot x^\sigma}{|\dot x|^2} 
+O(\varkappa^3)\,. 
\ee 
 
Thus, the anomalous acceleration is 
\be 
A^\mu{}_{\rm anom}=- 
\frac{\varkappa^2}{8}g^{\mu\nu}\left( 
2\nabla_\alpha S_{\beta\nu\rho\sigma} 
-\nabla_\nu S_{\alpha\beta\rho\sigma} 
\right) 
\frac{\dot x^\rho\dot x^\sigma\dot x^\alpha\dot x^\beta}{|\dot x|^2} 
+O(\varkappa^3)\,, 
\ee 
Notice that with our choice of the parameter $t$ we have 
$F(x,\xi(x,\xi))=1$, 
and, therefore, in the equations of motion we can substitute with 
the same accuracy 
\be 
|\xi|^2=-1+O(\varkappa^2)\,,\qquad 
|\dot x|^2=-1+O(\varkappa^2)\,. 
\ee 
Therefore, we obtain finally 
\be 
A^\mu{}_{\rm anom}=- 
\frac{\varkappa^2}{8}g^{\mu\nu}\left( 
2\nabla_\alpha S_{\beta\nu\rho\sigma} 
-\nabla_\nu S_{\alpha\beta\rho\sigma} 
\right) 
\dot x^\rho\dot x^\sigma\dot x^\alpha\dot x^\beta 
+O(\varkappa^3)\,. 
\label{283xx} 
\ee 
 
\subsection{Non-commutative Corrections to Newton's Law} 

Now, we will derive the non-commutative corrections to the Newton's Law.
We label the coordinates as 
\be 
x^0=t, \qquad x^1=r, \qquad x^2=\theta,\quad x^3=\varphi\,,
\ee 
and consider the static spherically symmetric (Schwarzschild) 
metric 
\be 
ds^2=-U(r)dt^2+U^{-1}(r)dr^2+r^2(d\theta^2+\sin^2\theta\,d\varphi^2)\,, 
\ee 
where 
\be 
U(r)=1-\frac{r_g}{r}\,,\qquad 
r_g=2GM\,, 
\ee 
and $M$ is the mass of the central body. 
It is worth recalling that here $t$ is the coordinate time. In the 
previous sections we used $t$ to denote an affine parameter of the 
trajectory that we agreed to choose to be the proper time. In the 
present section we use $\tau$ to denote the proper time and $t$ to 
denote the coordinate time. 

The motion of test particles
in Schwarzschild geometry is 
very well studied in General Relativity, see, for example 
\cite{weinberg72}. 
Assuming that the particle moves in the 
equatorial plane $\theta=\pi/2$ away from the center, that is, 
$dr/d\tau>0$, 
the equations of motion have the following integrals 
\cite{weinberg72} 
\bea 
\dot x^0=\frac{ dt}{d\tau}&=&\frac{E}{m}\frac{1}{U(r)}\,,\\ 
\dot x^1=\frac{dr}{d\tau}&=&\sqrt{\frac{E^2}{m^2} 
-\left(1+\frac{L^2}{m^2}\frac{1}{r^2}\right)U(r)}\,,\\ 
\dot x^2=\frac{d\theta}{d\tau}&=&0\,,\qquad 
\theta=\frac{\pi}{2}\,,\\ 
\dot x^3=\frac{d\varphi}{d\tau}&=&\frac{L}{m}\frac{1}{r^2}\,,\\ 
\eea 
where $m$, $L$, and $E$ are the mass of the particle, its orbital 
momentum and the energy. 

In the non-relativistic limit for weak gravitational fields, 
assuming 
\be 
E=m+E'\,, 
\ee 
with $E'<<m$, and $r>>r_g$
one can identify the coordinate time with the proper time, 
so that 
\be 
\dot x^0=\frac{dt}{d\tau}=1\,. 
\ee 
Further, for the non-relativistic motion we have
$\dot r$, $r\dot \theta$, $r\dot\varphi<<1$, 
and the radial velocity reduces, of course, to the standard Newtonian 
expression 
\bea 
\dot x^1=\frac{dr}{d\tau}&=&\sqrt{\frac{2E'}{m} 
-\frac{L^2}{m^2}\frac{1}{r^2} 
+\frac{r_g}{r}} 
\,,
\eea 
which for $L=0$ becomes
\bea 
\dot x^1=\frac{dr}{d\tau}&=&\sqrt{\frac{2E'}{m}
+\frac{r_g}{r}} 
\,,
\eea 

It is worth stressing that the anomalous acceleration due to 
non-commutativity in the non-relativistic limit
can be interpreted as a correction to the Newton's Law. Assuming that
a particle is moving in the equatorial plane, $\theta=\pi/2$,
with zero orbital momentum, $\varphi={\rm const}$, the equation of motion
is
\bea
\frac{d^2 r}{dt^2}&=&-\frac{\partial}{\partial r}V_{\rm eff}(r)
\nonumber\\
&=&-\frac{GM}{r^2}+A^r{}_{\rm anom}\,,
\eea
where in the uniform model
\be 
A^{r}{}_{\rm anom}= \frac{\varkappa^{2}}{8} \partial_r S^{0000} 
+O(\varkappa^3) 
\,,
\label{314xx}
\ee
with $S^{0000}=\frac{1}{N}\tr h^{00}h^{00}$,
and 
in the non-uniform model
\be 
A^{r}{}_{\rm anom}= -\frac{\varkappa}{2} \partial_r 
q^{00} 
+O(\varkappa^{2}) 
\label{315xx}
\,, 
\ee 
with $q^{00}$ being the component of the tensor $q^{\mu\nu}$ 
defined by (\ref{275xx}). 
This gives the non-commutative corrections to Newton's Law:
in the uniform model,
\bea
V_{\rm eff}(r)=-\frac{GM}{r}-\frac{\varkappa^{2}}{8} S^{0000}(r)
+O(\varkappa^3)\,,
\eea
and, in the nonuniform model,
\bea
V_{\rm eff}(r)=-\frac{GM}{r}+\frac{\varkappa}{2} q^{00}(r)
+O(\varkappa^2)
\,.
\eea
Here, of course, the tensor components $S^{0000}$ and $q^{00}$
should be obtained by the solution of the non-commutative
Einstein field equations (in the perturbation theory).

%
\section{Noncommutative Einstein Equations} 
\setcounter{equation}{0} 
 
The dynamics of the tensor field $a^{\mu\nu}$ is described by the 
action functional of Matrix Gravity. As was outlined in the 
Introduction, there is no unique way to construct such an action 
functional: there are at least two approaches, one 
\cite{avramidi04a}  based on a non-commutative generalization of 
Riemannian geometry and another one \cite{avramidi04b} based on 
the spectral asymptotics of a non-Laplace type partial 
differential operator.  
The exact 
non-commutative Einstein equations for 
the action functional proposed in \cite{avramidi04a} were found in 
our recent work \cite{fucci07}. 
The equations of motion for the spectral approach were obtained
within the perturbation theory in our paper
\cite{fucci08}.
In the present work we are using 
the approach of \cite{avramidi04a}. 
We should mention that in the perturbation theory 
the difference between these two approaches consists in just some 
numerical parameters of the action; the general structure of the 
terms is the same. 

In the following we will give a very brief 
overview of  the general formalism, more details can be found in 
\cite{avramidi04a, fucci07}. 
We define the matrix-valued tensor 
$b_{\mu\nu}$ by 
\be 
a^{\mu\nu}b_{\nu\lambda}=\delta^\mu_\lambda\II\,, 
\ee 
the matrix-valued connection coefficients 
${\cal A}^{\mu}{}_{\alpha\beta}$ by 
\begin{equation}\label{15f} 
{\cal A}^{\alpha}{}_{\lambda\mu} 
=\frac{1}{2}b_{\lambda\sigma}(a^{\alpha\gamma} 
\partial_{\gamma}a^{\rho\sigma} 
-a^{\rho\gamma}\partial_{\gamma}a^{\sigma\alpha} 
-a^{\sigma\gamma}\partial_{\gamma}a^{\alpha\rho})b_{\rho\mu}\;, 
\end{equation} 
and the matrix-valued Riemann curvature tensor 
\begin{equation} 
\label{4xxx} 
\mathcal{R}^{\lambda}{}_{\alpha\mu\nu} 
=\partial_{\mu}{\cal A}^{\lambda}{}_{\alpha\nu} 
-\partial_{\nu}{\cal A}^{\lambda}{}_{\alpha\mu} 
+{\cal A}^{\lambda}{}_{\beta\mu}{\cal A}^{\beta}{}_{\alpha\nu} 
-{\cal A}^{\lambda}{}_{\beta\nu}{\cal A}^{\beta}{}_{\alpha\mu}\;. 
\end{equation} 
Next, we define a matrix-valued density 
\begin{equation}\label{7xxx} 
\rho=\int\limits_{\mathbb{R}^{n}}\frac{d\xi}{\pi^{\frac{n}{2}}} 
\exp[{-a^{\mu\nu}\xi_{\mu}\xi_{\nu}}]\;. 
\end{equation} 
This enables us to define the  action of Matrix Gravity as follows 
\begin{equation}\label{8xxx} 
S_{\textrm{MG}}(a)=\frac{1}{16\pi 
G}\int\limits_{M}dx\; 
{\rm Re}\; 
\frac{1}{N}\tr\rho\;(a^{\mu\nu}\mathcal{R}^\alpha{}_{\mu\alpha\nu} 
-2\Lambda)\;, 
\end{equation} 
where $G$ is the Newtonian gravitational constant and $\Lambda$ is the 
cosmological constant. 
 
Of course, in the commutative limit all these constructions become 
the standard geometric background of General Relativity. The 
tensors $a^{\mu\nu}$ and $b_{\mu\nu}$ become the contravariant and 
covariant Riemannian metrics, the coefficients  ${\cal 
A}^\alpha{}_{\mu\nu}$ become the Christoffel symbols, the tensor 
${\cal R}^\alpha{}_{\beta\mu\nu}$ becomes the standard Riemann 
tensor and the action of Matrix Gravity becomes the Einstein 
action functional. In the presence of matter one should add to 
this functional the action of the matter fields and particles. The 
non-commutative Einstein equations were obtained in 
\cite{fucci07}. These equations, in full generality, are a 
complicated system of non-linear second-order partial differential 
equations. Their study is just beginning. Of course, it would be 
extremely interesting to obtain some simple exact solutions. 
 
In the present paper we study the effects 
of these equations in the simplest possible case restricting ourselves 
to a {\it commutative algebra}. 
The commutativity assumption enormously simplifies the dynamical 
equations. In this case they 
look exactly as the Einstein equations in the vacuum 
\begin{equation} 
\label{44} 
\mathcal{R}_{\mu\nu}=\Lambda b_{\mu\nu}\;, 
\end{equation} 
where ${\cal R}_{\mu\nu}$ is the matrix-valued Ricci tensor defined by 
\be 
{\cal R}_{\mu\nu}={\cal R}^\alpha{}_{\mu\alpha\nu}\,. 
\ee

\subsection{Static Spherically Symmetric Solutions} 
 
We study, now, the static spherically symmetric solution of the equation 
(\ref{44}).
We present the 
matrix-valued metric $a^{\mu\nu}$ by writing the ``matrix-valued 
Hamiltonian''
\begin{equation}\label{19xx} 
a^{\mu\nu}\xi_\mu\xi_\nu= 
A(r)(\xi_0)^2
+B(r)(\xi_1)^2 
+\mathbb{I}\frac{1}{r^2}
\left[(\xi_2)^2
+\frac{1}{\sin^{2}\theta}(\xi_3)^2\right]\;, 
\end{equation} 
or the ``matrix-valued interval''
\begin{equation}\label{19} 
b_{\mu\nu}dx^\mu dx^\nu= 
A^{-1}(r)dt^{2}+B^{-1}(r)dr^{2} 
+\mathbb{I}\;r^{2}\left(d\theta^{2} 
+\sin^{2}\theta\, d\varphi^{2}\right)\;, 
\end{equation} 
where the coefficients $A(r)$ and $B(r)$ 
are commuting matrices that depend only on the 
radial coordinate $r$. 
This simply means that we choose the following ansatz 
\begin{eqnarray}\label{29xxx} 
a^{00}&=& A\;,\qquad 
a^{11}=B\;, \nonumber\\[10pt] 
a^{22}&=&\frac{1}{r^2}\;\mathbb{I}\;, \qquad 
a^{33}=\frac{1}{r^{2}\sin^{2}\theta} \;\mathbb{I}\;. 
\end{eqnarray} 
 
Next, by computing the connection coefficients ${\cal A}^\alpha{}_{\mu\nu}$ and 
the matrix-valued Ricci tensor we obtain 
the equations 
of motion
\begin{eqnarray} 
\mathcal{R}_{00}&=&
A^{-1}B\left[
\frac{1}{2}A^{-1}A^{\prime\prime} 
-\frac{3}{4}A^{-2}(A^{\prime})^2 
+\frac{1}{4}A^{-1}A^{\prime}B^{-1}B^{\prime} 
+\frac{1}{r}A^{-1}A^{\prime}
\right] 
=\Lambda A^{-1}\;,
\nonumber
\\
\label{40}\\ 
\mathcal{R}_{11}&=& 
\frac{1}{2}A^{-1}A^{\prime\prime} 
-\frac{3}{4}A^{-2}(A^{\prime})^2 
+\frac{1}{4}A^{-1}A^{\prime}B^{-1}B^{\prime}
-\frac{1}{r}B^{-1}B^{\prime}
=\Lambda B^{-1}\;, 
\label{41}\\ 
\mathcal{R}_{22}&=& 
-\frac{r}{2}B'
-B 
+\frac{r}{2}BA^{-1}A^{\prime}
+\mathbb{I}
=\Lambda 
r^2\cdot\II\;, 
\label{42}\\ 
\mathcal{R}_{33}&=&\sin^{2}\theta\;\mathcal{R}_{22} =\Lambda 
r^2\sin^2\theta\cdot\II \;. \label{43} 
\end{eqnarray} 
where the prime 
denotes differentiation with respect to $r$. 
 
By using the equations (\ref{40}) and (\ref{41}) we find 
\begin{equation}\label{46} 
A^{-1}A^{\prime}+B^{-1}B^{\prime}=0\;; 
\end{equation} 
the general solution of this equation is 
\begin{equation}\label{47} 
A(r)B(r)=C_1\;, 
\end{equation} 
where $C_1$ is an arbitrary constant matrix from our algebra. 
We require that at the spatial infinity as $r\to \infty$ the matrices 
$A$ and $B$ and, therefore, the matrix $C$ as well, 
are non-degenerate.

By using 
this relation we obtain further  from eqs. (\ref{41}) 
and (\ref{42}) two 
compatible equations for the matrix $B$ 
\be 
B''+\frac{2}{r}B'+2\Lambda=0\,, 
\label{40xxx} 
\ee 
and 
\be 
rB'+B=(1-\Lambda r^2)\II\,. 
\label{42xxx} 
\ee 
The general solution of the eq. (\ref{42xxx}) is 
\be 
B(r)=\left(1-\frac{1}{3}\Lambda r^2\right)\II+\frac{1}{r}C_2\,, 
\ee 
where $C_2$ is another arbitrary constant matrix  from our algebra. It 
is not difficult to see that this form of the matrix $B$ also satisfies 
the eq. (\ref{40xxx}). The matrix $A$ is now obtained from the equation 
(\ref{47}) 
\be 
A(r)=C_1\left[\left(1-\frac{1}{3}\Lambda r^2\right)\II 
+\frac{1}{r}C_2\right]^{-1}\,. 
\ee 

We will also 
require that in the limit 
$\varkappa\rightarrow 0$ we 
should get the standard 
Schwarzschild solution with the cosmological constant
\be 
B(r)=-A^{-1}(r)=\left(1-\frac{1}{3}\Lambda r^2 
-\frac{r_g}{r}\right)\II\,, 
\ee 
where $r_g$ is the gravitational radius of the central body of mass
$M$,
\be 
r_g=2GM\,,
\label{330xx}
\ee 
that is, in that limit the matrices $C_1$ and $C_2$ should be
\be
C_1=-\II\,, \qquad C_2=-r_g\II\,.
\ee
 
\subsection{$2\times 2$ Matrices} 
 
To be specific, we restrict ourselves further 
to real symmetric $2\times 2$ matrices 
generated by 
\begin{equation}\label{13a} 
\mathbb{I}= 
\left(%
\begin{array}{cc} 
  1 & 0 \\ 
  0 & 1 \\ 
\end{array}%
\right) 
\,,\qquad 
\textrm{and}\qquad\tau=\left(%
\begin{array}{cc} 
  0 & 1 \\ 
  1 & 0 \\ 
\end{array}%
\right)\;. 
\end{equation} 
In this case the 
constant matrices $C_1$ and $C_2$ can be expressed in terms of four 
real parameters 
\be 
C_1=\alpha\II+\theta\tau\,,\qquad 
C_2=\mu\II+L\tau\,, 
\ee 
where $\theta=\varkappa\bar\theta$ and $L=\varkappa\bar L$ are the 
parameters of first order in the deformation parameter $\varkappa$.
Here the parameters $\alpha$ and $\theta$ are dimensionless and 
the parameters $\mu$ and $L$ have the dimension of length. 

Then the matrix 
$B(r)$ has the form 
\begin{equation}\label{20} 
B(r)=\left(1-\frac{1}{3}\Lambda r^2+\frac{\mu}{r}\right)\II
+\frac{L}{r}\tau\,.
\end{equation} 
Next, noting that $\tau^2=\II$, and by using the relation
\be 
(a\II+b\tau)^{-1}=\frac{1}{a^2-b^2}\left(a\II-b\tau\right)\,, 
\ee 
we obtain the matrix $A(r)$ 
\be 
A(r)=\varphi(r)\II+\psi(r)\tau\,,
\ee 
where 
\be 
\varphi(r)=\frac{\alpha\left(1-\frac{1}{3}\Lambda r^2\right)
+\frac{\alpha\mu-\theta L}{r}}
{\left(1-\frac{1}{3}\Lambda r^2+\frac{\mu}{r}\right)^2
-\frac{L^2}{r^2}}\,, 
\ee 

\be 
\psi(r)=\frac{\theta\left(1-\frac{1}{3}\Lambda r^2\right)
+\frac{\theta \mu-\alpha L}{r}}{\left(1-\frac{1}{3}\Lambda r^2 
+\frac{\mu}{r}\right)^2 
-\frac{L^2}{r^2}}\,. 
\ee 
 
The parameters $\alpha, \theta, \mu$ and $L$ should be determined 
by the boundary conditions at spatial infinity. The question of 
boundary conditions is a subtle point since we do not know the 
physical nature of the additional degrees of freedom. 
We will simply require that the diagonal part of the metric
is asymptotically De Sitter. This
immediately gives
\be 
\alpha=-1\,.
\ee 
Now, we introduce a new parameter
\be
r_0=|\Lambda|^{-1/2}\,,
\ee
and require that for $r_g<<r<<r_0$, the diagonal part of the
metric, more precisely, the function $\varphi(r)$ is
asymptotically Schwarzschild, that is,
\be
\varphi(r)=-1-\frac{r_g}{r}+O\left(\frac{r_g^2}{r^2}\right)
+O\left(\frac{r^2}{r_0^2}\right)\,.
\ee
This fixes the parameter $\mu$
\be 
\mu=-r_g+\theta L\,. 
\ee 
The parameters $\theta$ and $L$ remain undetermined. 
 
Finally, by introducing new parameters
\be 
\rho=(1+\theta^2) L -\theta r_g 
\ee 
\be 
r_{\pm}=r_g-(\theta\pm 1)L\, 
\ee 
we can rewrite our solution in the form
\be 
\varphi(r)=\frac{-r\left(r-\frac{1}{3}\Lambda r^3
-r_g+2\theta L\right)}
{\left[r-\frac{1}{3}\Lambda r^3-r_-\right]
\left[r-\frac{1}{3}\Lambda r^3-r_+\right]}\,, 
\ee 

\be 
\psi(r)=\frac{r\left[\theta\left(r-\frac{1}{3}\Lambda r^3\right)
+\rho\right]}
{\left[r-\frac{1}{3}\Lambda r^3-r_-\right]
\left[r-\frac{1}{3}\Lambda r^3-r_+\right]}\,
\label{3125xx}
\,. 
\ee 
Of course, as $\varkappa\rightarrow 0$
both parameters $L=\varkappa\bar L$ and 
$\theta=\varkappa\bar\theta$ 
vanish and we 
get the standard 
Schwarzschild solution with the cosmological constant.

Notice that the matrix-valued metric $a^{\mu\nu}$ 
becomes singular when 
the matrices $A$ and $B$ are not invertible, that is, 
when 
\be 
\det A(r)=0\,. 
\ee 
The solutions of this equation are the roots of the cubic 
polynomials 
\be 
r-\frac{1}{3}\Lambda r^3-r_-=0 
\qquad\mbox{and}\qquad 
r-\frac{1}{3}\Lambda r^3-r_+=0 
\label{4146} 
\ee 
 
Recall that the standard Schwarzschild coordinate singularity,  which 
determines the position of the event horizon, is located at $r=r_g$. The 
presence of singularities depends on the values of the parameters. We 
analyze, now, the first eq. in (\ref{4146}). In the case $\Lambda\le 0$ 
the polynomial has one root if $r_->0$ and does not have any roots if 
$r_-<0$. In the case $\Lambda>0$ it is easy to see that: i) if $r_- 
>(2/3) r_0$, then there are no roots, ii) if $0< r_- <(2/3) r_0$, then 
the polynomial has two roots, and ii) if $r_-<0$, then the polynomial 
has one root. The same applies to the second eq. in (\ref{4146}). 
 
We emphasize that there are two cases {\it without any singularities at 
any finite value of $r$}. This happens if either: a) $\Lambda\le 0$ and 
$r_\pm <0$, or b) $\Lambda>0$ and $r_\pm> (2/3)r_0$. This can certainly 
happen for large values of $|\theta|$ and $|L|$. 
In particular, if 
$\theta$ and $L$ have the same signs and 
\be 
|\theta|>1+\frac{r_g}{|L|}\,, 
\ee 
then both $r_\pm$ are negative, $r_\pm <0$, and 
if $\theta$ and $L$ have opposite 
signs and 
\be 
|\theta|>1+\frac{\frac{2}{3}r_0-r_g}{|L|}\,, 
\ee 
then $r_\pm> (2/3)r_0$. This is a very interesting phenomenon which is 
entirely new and due to the additional degrees of freedom. 
 
We would like to clarify some points. 
The parameters $\mu_{i}$ introduced in 
the previous sections describe the properties of 
the test particle, that is, the matter. 
The parameters  
$\theta$ and $\rho$ introduced 
in the static and spherically symmetric solution of non-commutative Einstein 
equations describe the properties of the gravitational field, that is,
the properties of the source of the gravitational field, that is, the
central body.
The parameters $\theta$ and $\rho$ are not related to the parameters $\mu_i$.

\section{Anomalous Acceleration} 
\setcounter{equation}{0}

In 
this section we are going to evaluate the anomalous acceleration 
of non-relativistic test particles in the static spherically symmetric
gravitational field of a massive central body. 

All we have to do is to evaluate the 
components of the anomalous acceleration (\ref{283xx}). 
As we will see the only essential component of the anomalous 
acceleration is the radial one $A^r{}_{\rm anom}$. 
All other components of 
the anomalous acceleration are negligible in this limit. 
As we will see below, the anomalous acceleration is caused by 
the radial gradient of the 
component $h^{00}$ of the matrix-valued metric, which is
\be 
\varkappa h^{00}
=\psi(r)\tau\,, 
\label{5158} 
\ee 
where $\psi(r)$ is given by 
(\ref{3125xx}). 
Our analysis is restricted 
to the perturbation theory in the deformation parameter 
$\varkappa$ (first order in $\varkappa$ in the non-uniform model 
and second order in $\varkappa$ in the uniform model). 
That is, we should expand our result in powers of $\rho$ and $\theta$ and 
keep only linear terms in the non-uniform model and quadratic  
terms in the uniform model. 

For future use we write the function $\psi(r)$
in the first order in the parameter $\varkappa$
\be 
\psi(r)=\frac{r\left[\theta\left(r-\frac{1}{3}\Lambda r^3\right)
+\rho\right]}
{\left(r-\frac{1}{3}\Lambda r^3-r_g\right)^2}
+O(\varkappa^2)\,
\,, 
\ee 
and for $r<<r_0$ 
\be 
\psi(r)=\frac{r(\theta r
+\rho)}{(r-r_g)^2}
+O(\varkappa^2)
\,,
\label{3125xxz}
\ee 
and, finally, for $r_g<<r<<r_0$,
\be 
\psi(r)=\theta+\frac{\rho}{r}
+O(\varkappa^2)
\,\,.
\label{3125xxzt}
\ee

We would like to emphasize at this point 
that the perturbation theory we are going to perform is only 
valid for small corrections. When the corrections become large
we need to consider the exact equations of motion
(\ref{12}). 

\subsection{Uniform Model} 
 
In the non-relativistic limit the formula 
for the anomalous radial acceleration 
(\ref{314xx}) gives
\bea
\label{333zz} 
 A^{r}{}_{\rm anom} 
&=& \frac{1}{4}\psi(r)\psi'(r)
+O(\varkappa^3)\,.
\eea 
The derivative of the function $\psi(r)$ is easily computed
\be
\psi'(r)=\omega(r)\psi(r)\,,
\ee
where
\be
\omega(r)=
\frac{1}{r}
+\frac{\theta(1-\Lambda r^2)}
{\theta\left(r-\frac{1}{3}\Lambda r^3\right)+\rho}
-\frac{1-\Lambda r^2}
{r-\frac{1}{3}\Lambda r^3-r_-}
-\frac{1-\Lambda r^2}
{r-\frac{1}{3}\Lambda r^3-r_+}\,.
\ee
Thus, we obtain finally
\bea
\label{333yy} 
 A^{r}{}_{\rm anom} 
&=& \frac{1}{4}\psi^2(r)\omega(r)
+O(\varkappa^3)\,.
\eea 

Recall that the parameters $\rho$ and $\theta$ are of first order 
in $\varkappa$.
Strictly speaking we should expand this formula 
in $\rho$ and $\theta$ keeping only quadratic terms;
we get 
\bea
A^{r}{}_{\rm anom}
&=&\frac{1}{4}
\frac{\left[\theta \left(r-\frac{1}{3}\Lambda r^3\right)+\rho\right]r}
{\left(r-\frac{1}{3}\Lambda r^3-r_g\right)^5}
\Biggl\{
\left(r-\frac{1}{3}\Lambda r^3-r_g\right)
\left[\theta \left(2r-\frac{4}{3}\Lambda r^3\right)+\rho\right]
\nonumber\\
&&
-2r(1-\Lambda r^2)
\left[\theta \left(r-\frac{1}{3}\Lambda r^3\right)+\rho\right]
\Biggr\}
+O(\varkappa^3)\,.
\eea
For $r<<r_0$ (that is, $|\Lambda| r^2<<1$) this becomes
\bea
A^{r}{}_{\rm anom}
&=&-\frac{1}{4}
\frac{r\left(\theta r+\rho\right)
\left[(\rho+2\theta r_g)r+\rho r_g-\frac{2}{3}\theta\Lambda r^4\right]}
{(r-r_g)^5}
+O(\varkappa^3)\,.
\eea
We need to keep the term linear in $\Lambda$ since we do not know
the values of the parameters $\theta$ and $\rho$.
Finally, for $r_g<<r<<r_0$ we obtain
\bea
A^{r}{}_{\rm anom}
&=&-\frac{1}{4}
\left(\theta + \frac{\rho}{r}\right)
\left(\frac{\rho+2\theta r_g}{r^2}
-\frac{2}{3}\theta\Lambda r\right)
+O(\varkappa^3)\,.
\eea

\subsection{Non-uniform Model} 
 
Similarly, in the non-uniform model the anomalous acceleration is
given by
eq. (\ref{315xx}).
In the $2\times 2$ matrix case considered above 
the eigenvalues of the matrix 
$ 
h^{\mu\nu}\xi_\mu\xi_\nu 
$ 
are 
\be 
\lambda_{1,2}=\pm \frac{1}{2}\tr( h^{\mu\nu}\tau)\xi_\mu\xi_\nu\,. 
\ee 
Therefore, 
\be 
P(x,\xi)=\mu_1\lambda_1+\mu_2\lambda_2 
=\gamma \frac{1}{2}\tr( h^{\mu\nu}\tau)\xi_\mu\xi_\nu\,, 
\ee 
where 
\be 
\gamma=\mu_1-\mu_2\,. 
\ee 
Thus 
\be 
q^{\mu\nu}=\frac{\gamma}{2}\tr( h^{\mu\nu}\tau)\,. 
\ee 
So, we 
obtain 
\be 
\varkappa q^{00}=\gamma\psi(r)\,. 
\ee 
Thus
\bea
\label{222zz} 
A^{r}_{\rm anom}
&=&-\frac{1}{2}\gamma\psi'(r)+O(\varkappa^2)
\nonumber\\
&=&-\frac{1\gamma}{2}\psi(r)\omega(r)
+O(\varkappa^2)\,.
\eea

Now, we recall that $\rho$ and $\theta$ are of first order in $\varkappa$
and expand in powers of $\rho$ and $\theta$ 
keeping only linear terms 
\bea
\label{222xzz} 
A^{r}_{\rm anom}
&=&-\frac{1}{2}\frac{\gamma}
{\left(r-\frac{1}{3}\Lambda r^3-r_g\right)^3}
\Biggl\{\left(r-\frac{1}{3}\Lambda r^3-r_g\right)
\left[\theta\left(2r-\frac{4}{3}\Lambda r^3\right)+\rho\right]
\nonumber\\
&&
-2r(1-\Lambda r^2)\left[
\theta\left(r-\frac{1}{3}\Lambda r^3\right)+\rho
\right]\Biggr\}
+O(\varkappa^2)\,.
\eea
In the case $r<<r_0$ (when $|\Lambda| r^2<<1$) this takes the form
\bea
\label{222xzza} 
A^{r}_{\rm anom}
&=&\frac{1}{2}\gamma
\frac{\left[(\rho+2\theta r_g)r+\rho r_g
-\frac{2}{3}\theta\Lambda r^4\right]}
{(r-r_g)^3}
+O(\varkappa^2)\,.
\eea
Finally, for $r_g<<r<<r_0$ we obtain
\bea
\label{222xzzb} 
A^{r}_{\rm anom}
&=&\frac{1}{2}\gamma
\left[\frac{(\rho+2\theta r_g)}{r^2}
-\frac{2}{3}\theta\Lambda r
\right]
+O(\varkappa^2)\,.
\eea

\section{Conclusions} 
 
In this paper we described the kinematics of test particles in the 
framework of a recently developed modified theory of gravitation,
called Matrix Gravity \cite{avramidi03,avramidi04a,avramidi04b}. 
We outlined the motivation for this theory, which is a non-commutative
deformation of General Relativity. Matrix Gravity can be interpreted in terms
of a collection of Finsler geometries on the spacetime manifold rather than 
in terms of Riemannian geometry.
This leads, in particular, to a 
new phenomenon of \emph{splitting} of Riemannian geodesics into a 
system of trajectories (Finsler geodesics) close to the Riemannian
geodesic. More precisely, instead of one 
Riemannian metric we have several Finsler metrics and different 
mass parameters which describe the tendency to follow a particular 
Finsler geodesics determined by a particular Finsler metric.
As a result the test particles 
exhibit a \emph{non-geodesic motion} which can be interpreted in terms of 
an anomalous acceleration.

By using a commutative algebra we found 
a static spherically symmetric solution of the 
modified Einstein equations. 
In this case a 
completely new feature appears due to the presence of 
additional degrees of freedom. The coordinate singularities of our 
model depend of additional parameters (constants of integration). 
Interestingly,
there is a  range of values for these free parameters in which 
\emph{no singularity occurs}. This is just one of the intriguing 
differences between Matrix Gravity and General Relativity. 

The description of matter in Matrix Gravity needs additional study.
In this paper we studied just the behavior of classical test particles.
We propose to describe a gravitating particle by several mass parameters rather
than one parameter as in General Relativity.
We considered two models of matter: a uniform one, in which all mass parameters are
equal, and a non-uniform one, in which the mass parameters are different.
The choice of one model over the other should be
dictated by physical reasons. 
It is worth emphasizing that in the generic non-uniform model
the {\it equivalence principle is violated}.

The interesting question 
whether the matter is described by only one mass 
parameter or more than one mass parameters as well as the more general question
of the physical origin of multiple mass
parameters requires further study. 
Since we do not know much about the physical  origin of these
mass parameters masses, 
we do not have to assume that they are positive. We do not exclude the
possibility that some of the mass parameters can be negative or zero.
This would imply, of course, that in this theory there is also
gravitational repulsion (antigravity). This could help solve the problem
of the gravitational collapse in General Relativity, which is caused by
the  infinite gravitational attraction. 

The next step of our analysis of the phenomenological 
consequences of Matrix Gravity is to apply the kinematic model developed 
in the previous sections to the study of such effects as the motion of 
Pioneer spacecrafts (Pioneer anomaly) and galactic rotations (dark matter). 
It would be 
very interesting to understand if the 
anomalous acceleration of the spacecrafts and the
flat rotation curves of galaxies 
can be explained without the concept of dark matter. 
 

\end{document}